\documentclass[aps,preprintnumbers,superscriptaddress,nofootinbib]{revtex4}

\usepackage{graphicx}
\def\beq{\begin{equation}}
\def\eeq{\end{equation}}
\def\bea{\begin{eqnarray}}
\def\eea{\end{eqnarray}}

\def\gappeq{\mathrel{\rlap {\raise.5ex\hbox{$>$}}
{\lower.5ex\hbox{$\sim$}}}}

\def\lappeq{\mathrel{\rlap{\raise.5ex\hbox{$<$}}
{\lower.5ex\hbox{$\sim$}}}}

\def\ga{\mathrel{\raise.3ex\hbox{$>$\kern-.75em\lower1ex\hbox{$\sim$}}}}
\def\la{\mathrel{\raise.3ex\hbox{$<$\kern-.75em\lower1ex\hbox{$\sim$}}}}
\def\gev{{\rm \, Ge\kern-0.125em V}}
\def\tev{{\rm \, Te\kern-0.125em V}}
\def\beq{\begin{equation}}
\def\eeq{\end{equation}}

\def\ohsq{\Omega_{\chi} h^2}

\def\m12{m_{1\!/2}}

\begin{document}
\bibliographystyle{revtex}

\preprint{CERN--TH/2001--340, UMN--TH--2034/01, TPI--MINN--01/53, 
hep-ph/0112013, Snowmass P3--47}

\title{Post-LEP CMSSM Benchmarks for Supersymmetry}
\author{M.~Battaglia}
\affiliation{CERN, CH-1211 Geneva 23, Switzerland}
\author{A.~De Roeck}
\affiliation{CERN, CH-1211 Geneva 23, Switzerland}
\author{J.~Ellis}
\affiliation{CERN, CH-1211 Geneva 23, Switzerland}
\author{F.~Gianotti}
\affiliation{CERN, CH-1211 Geneva 23, Switzerland}
\author{K.~T.~Matchev}
\affiliation{CERN, CH-1211 Geneva 23, Switzerland}
\author{K.~Olive}
\affiliation{University of Minnesota, Minneapolis, MN 55455, USA}
\author{L.~Pape}
\affiliation{CERN, CH-1211 Geneva 23, Switzerland}
\author{G.~W.~Wilson}
\affiliation{University of Kansas, Lawrence, KS 66045, USA}
\date{December 2, 2001}

\begin{abstract}
\vspace*{0.25cm}

We introduce a set of CMSSM benchmark scenarios that take into account the
constraints from LEP, Tevatron, $b \rightarrow s \gamma$, $g_\mu - 2$ and
cosmology. The benchmark points are chosen to span the range of different
generic possibilities, including focus-point models, points where
coannihilation effects on the relic density are important, and points with
rapid relic annihilation via direct-channel Higgs poles, as well as points
with smaller sparticle masses. We make initial estimates of the physics
reaches of different accelerators, including the LHC, and $e^+ e^-$
colliders in the sub- and multi-TeV ranges. We stress the complementarity
of hadron and lepton colliders, with the latter favoured for
non-strongly-interacting particles and precision measurements.

\end{abstract}

\maketitle

\section{Introduction}

The completion of the LEP experimental programme has brought to an end an
era
of precise electroweak measurements and the search for new particles 
with masses $\lappeq 100$~GeV. With the start of Tevatron Run~II, the 
advent of the LHC and hopefully a linear $e^+ e^-$ collider, the 
experimental exploration of the TeV energy scale is beginning in earnest. 

The best-motivated scenario for new physics beyond  the Standard Model (SM) 
at the TeV energy scale is generally agreed to be Supersymmetry. 
Theoretically, it is compellingly
elegant, offers the possibility of unifying fermionic matter particles
with bosonic force particles, is the only framework thought to be
capable of connecting gravity with the other interactions, and appears
essential for the consistency of string theory. However, none of these
fundamental arguments offer clear advice as to the energy scale at which
supersymmetric particles might appear. Preserving the gauge hierarchy in a
natural way, however, motivates supersymmetry at the TeV scale.
Supersymmetry suggests the existence of a light Higgs boson, which is
favoured indirectly by precision electroweak data. 
If a Higgs particle weighing less than about 130 GeV
is discovered at the Tevatron, testing for the existence 
of supersymmetric particles and exploring their properties 
would become a prime focus of the experiments at the LHC.

As an aid to the comparative assessment of the prospects for 
detecting and measuring these sparticles
at different accelerators, benchmark sets of supersymmetric 
parameters have often been found useful, since they provide a focus for
concentrating the 
discussion~\cite{Hinchliffe:1997iu,unknown:1999fr,Aguilar-Saavedra:2001rg}. 
Here we review a recently-proposed set of post-LEP benchmark points that
take into account constraints derived from the direct searches for 
sparticles and Higgs bosons, the measurement of the
$b\rightarrow s\gamma$ branching ratio, and the preferred cosmological 
density range, within the framework of the 
constrained MSSM (CMSSM)~\cite{Battaglia:2001zp}. 
Some of our points have been adopted by the working groups at
Snowmass 2001 in defining the benchmark `Snowmass slopes'  1--4.
Input parameters in the CMSSM are universal gaugino masses 
$m_{1/2}$, scalar masses $m_0$ (including those of the Higgs multiplets) and
trilinear supersymmetry breaking parameters $A_0$ at the supersymmetric grand 
unification scale, together with  $\tan \beta$ and the sign of $\mu$.
This framework has the merit of being sufficiently specific that the
different phenomenological constraints can be combined meaningfully. On
the other hand, it is just one of the phenomenological possibilities offered by 
supersymmetry, and others also merit study. 

\section{Constraints}

Important constraints on the CMSSM parameter space are provided by direct
sparticle searches at LEP and the Tevatron collider. Also important is the
LEP limit on the Higgs mass $m_H>$114.1~GeV \cite{:2001xw}. 
This holds in the Standard
Model and, for the lightest Higgs boson $h$, in the general MSSM for
$\tan\beta \lappeq 8$ and for all $\tan\beta$ in the CMSSM cases of interest, 
at least as long as CP is conserved. This limit imposes important indirect constraints
on the CMSSM parameters, principally $m_{1/2}$. Finally, the loop-mediated
$b\rightarrow s\gamma$ transition is sensitive to chargino, squark and charged
Higgs masses. The $b\rightarrow s\gamma$ 
measurement~\cite{Ahmed:1999fh,Abe:2001hk} 
is currently compatible with the rate predicted in the SM, 
thus restricting the possible mass range of those superpartners. 
This constraint is more important for $\mu < 0$ but
is also significant for $\mu > 0$ when $\tan\beta$ is large. 

The cosmological constraints on the CMSSM are set by requiring that the
supersymmetric relic density $\rho_\chi = \Omega_\chi \rho_{critical}$
falls within the preferred range $0.1 < \Omega_\chi h^2 < 0.3$. The upper
limit is rigorous, since astrophysics and cosmology tell us that the total
matter density $\Omega_m \lappeq 0.4$, and the Hubble expansion rate $h
\sim 1/\sqrt{2}$ to within about 10\% (in units of 100 km/s/Mpc). On the other
hand, the lower limit is optional, since there could be additional
important contributions, other than sparticles, to the overall matter
density.  There are generic regions of the CMSSM parameter space where the
relic density falls within the preferred range. Since the relic density
typically increases with the relic mass, one might expect
an upper limit on the mass of the lightest superparticle (LSP) 
$m_\chi \lappeq$ 1 TeV. However, there are various ways in which 
this generic upper bound on $m_\chi$ can be evaded.  
For example, the relic density may be suppressed by
coannihilation
\cite{Griest:1991kh,Mizuta:1993qp,Edsjo:1997bg,Ellis:1998kh,Ellis:1999mm,%
Gomez:1999dk,Boehm:1999bj,Ellis:2001ms,Arnowitt:2001yh}
and the allowed CMSSM region may acquire a `tail' extending to large
$m_\chi$, as in the case where the next-to-lightest superpartner 
(NLSP) is the lighter stau, $\tilde\tau_1$, and $m_{\tilde\tau_1} \sim
m_\chi$
\cite{Ellis:1998kh,Ellis:1999mm,Gomez:1999dk,Arnowitt:2001yh}.  Another
mechanism is rapid annihilation via a direct-channel pole when $m_\chi
\sim {1\over 2} m_{Higgs, Z}$ \cite{Ellis:2001ms,Lahanas:2001yr}. This may
yield a `funnel' extending to large $m_{1/2}$ and $m_0$ at large
$\tan\beta$. Another allowed region at large $m_0$ is the `focus-point'
region~\cite{Feng:1999mn,Feng:1999zg,Feng:2000gh}, where the LSP has a
sizable higgsino component, enhancing its annihilation.

These filaments extending the preferred CMSSM parameter space are clearly
unconventional, but they cannot be excluded, and we think it important to
investigate the sensitivity of future planned and proposed colliders to
their phenomenology. 

\section{Proposed Benchmarks}

\begin{figure}[t]
\includegraphics[scale=0.48]{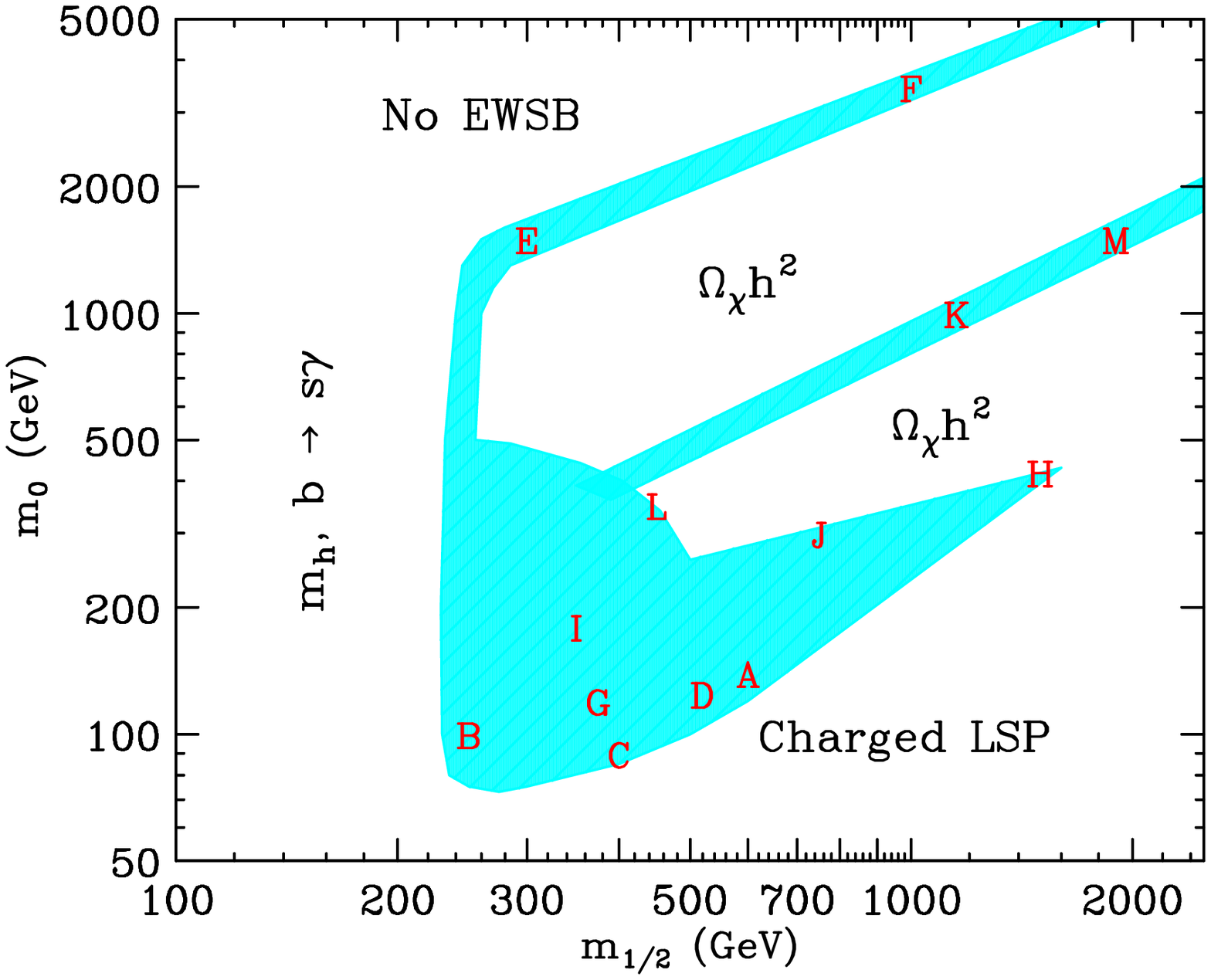}
\includegraphics[scale=0.48]{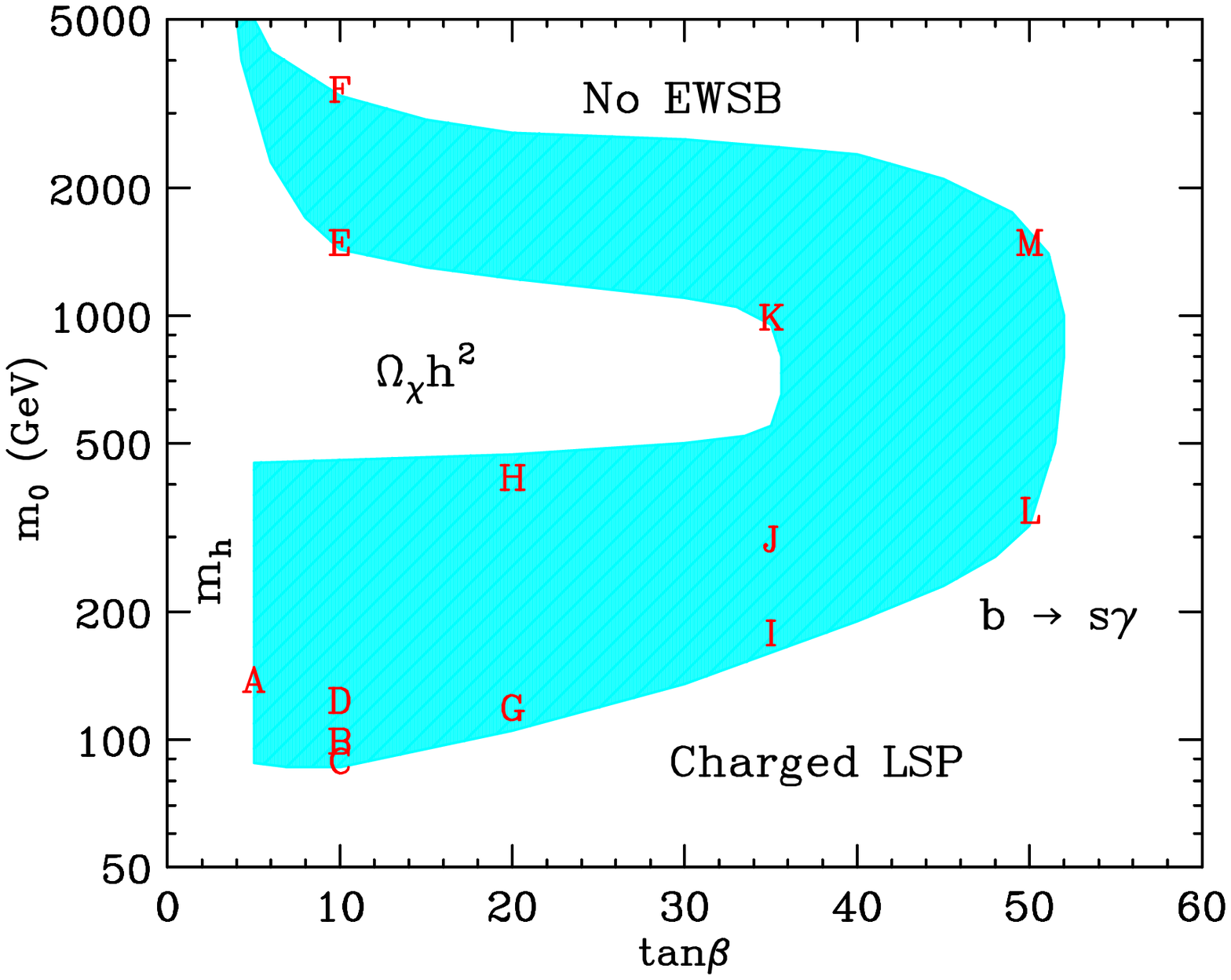}
\caption{Locations of our proposed CMSSM benchmark
points~\cite{Battaglia:2001zp} in 
(a) the $(m_{1/2}, m_0)$ plane, and (b) the $(\tan\beta, m_0)$ plane. 
The shaded areas roughly indicate the various cosmologically preferred
regions discussed in the text.}
\label{fig:1}
\end{figure}

The above constraints and limits define allowed regions in the $(m_{1/2}, m_0)$
plane which are qualitatively illustrated in Fig.~\ref{fig:1}a. 
Electroweak symmetry breaking (EWSB) is not possible in the top left corner, 
and the LSP would be charged in the bottom right region. 
The experimental constraints on $m_h$ and 
$b \rightarrow s \gamma$ exert pressures from the left,
depending on the exact value of $\tan \beta$ and the sign of $\mu$. 
In the remaining unshaded areas to the right the relic density is too
large and the Universe is overclosed. We observe a central (`bulk')
allowed region. The three filaments extending away from it are
(from top to bottom) the `focus-point' region, the rapid-annihilation
`funnel' and the coannihilation region.

In Fig.~\ref{fig:1}b we show the corresponding allowed regions
in the $(\tan\beta, m_0)$ plane. The absence of EWSB excludes the areas at 
the top and to the right (where $\mu^2<0$ and $m_A^2<0$, correspondingly).
The $m_h$ constraint is effective at low $\tan\beta$,
while the bottom area is ruled out because the LSP is charged.
The $b\to s \gamma$ constraint 
is maximally sensitive for large $\tan\beta$ and light superpartners, 
i.e., in the lower right corner. Finally, the relic density is too large 
in the remaining unshaded area in the middle.
One can still recognize three distinct areas inside the allowed region:
the `focus point' branch at the top, the vertical band on the right, due to
the rapid annihilation `funnel', and the horizontal band at the bottom, 
comprising the `bulk' and `coannihilation' regions.

Within these allowed domains of CMSSM parameter space,
thirteen benchmark points have been proposed, as sets of $m_{1/2}$, $m_0$, 
$\tan\beta$ and $sgn(\mu)$ values defining the entire spectrum of sparticles. 
These are given in Table~\ref{tab:1}, while the details of the corresponding 
spectra are to be found in~\cite{Battaglia:2001zp}. In order to reduce the
number of free parameters and in the absence of clear guidance from 
experimental and theory constraints, for simplicity we have set $A_0 = 0$.
Small nonzero values of $A_0$ have very little impact on phenomenology,
because of the fixed point structure of the $A$-term
renormalization-group equations. 
In order to obtain sufficiently distinct spectra, one must consider
rather large values of $A_0$. The inputs listed in the
Table have been used with the {\tt SSARD} programme to calculate the last
three lines. For the convenience of experimental simulations, 
in~\cite{Battaglia:2001zp} we have also provided inputs
for {\tt ISASUGRA 7.51} which reproduce the relevant features of the 
benchmark spectra as closely as possible.

\begin{table}[tbh]
\caption{The CMSSM parameters for the benchmark points proposed. 
In addition to the relic density $\ohsq$, the supersymmetric
contribution to $a_\mu \equiv (g_\mu - 2)/2$ (in units of $10^{-10}$), and
the $b \rightarrow s\gamma$ decay branching ratio (in units of $10^{-4}$) are given.} 
\begin{tabular}{|c||r|r|r|r|r|r|r|r|r|r|r|r|r|}
\hline
Model          & A   &  B  &  C  &  D  &  E  &  F  &  G  &  H  &  I  &  J  &  K  &  L  &  M   \\ 
\hline
$m_{1/2}$      & 600 & 250 & 400 & 525 &  300& 1000& 375 & 1500& 350 & 750 & 1150& 450 & 1900 \\
$m_0$          & 140 & 100 &  90 & 125 & 1500& 3450& 120 & 419 & 180 & 300 & 1000& 350 & 1500 \\
$\tan{\beta}$  & 5   & 10  & 10  & 10  & 10  & 10  & 20  & 20  & 35  & 35  & 35  & 50  & 50   \\
sign($\mu$)    & $+$ & $+$ & $+$ & $-$ & $+$ & $+$ & $+$ & $+$ & $+$ & $+$ & $-$ & $+$ & $+$  \\ 
\hline
$\ohsq$   & 0.26 & 0.18 & 0.14 & 0.19 & 0.31 & 0.17 & 0.16 & 0.29 & 0.16 &
0.20 & 0.19 & 0.21 &  0.17 \\
\hline
$\delta a_{\mu}$
 & 2.8 & 28 & 13 &-7.4 & 1.7 &0.29 & 27 & 1.7 & 45 & 11 &-3.3 & 31 &  2.1 \\
\hline
$B_{s \gamma}$
           &  $3.54$    &  $2.80$   &  $3.48$  
& $4.07$    &  $3.40$   & $3.32$    & $3.10$    & $3.28$ & $2.55$    & $3.21$   
& $ 3.78$   &  $2.71$   &  $ 3.24$  \\
\hline
\end{tabular}
\label{tab:1}
\end{table}

The recent precise measurement~\cite{Brown:2001mg}
of the anomalous magnetic moment of the muon, 
$g_\mu - 2$, which is in apparent disagreement with the SM at the $\simeq2.5\sigma$ 
level, can also be used to derive constraints on the CMSSM parameters
\cite{Feng:2001tr,Chattopadhyay:2001vx,Komine:2001fz,Ellis:2001yu,Arnowitt:2001be,%
Martin:2001st,Baer:2001kn,Komine:2001rm}. It disfavours 
$\mu < 0$ and large values of $m_0$ and $m_{1/2}$ for $\mu > 0$.
However, as the experimental accuracy is soon expected to be significantly 
improved and consensus on the calculation of hadronic contributions to 
$g_\mu - 2$ has yet be reached~\footnote{We note, in particular, the
current questioning of the sign of the light-by-light scattering
contribution~\cite{Knecht:2001qf,Knecht:2001qg}.}, we have chosen
not to apply strictly this constraint in the definition of the benchmarks
here. However, our choice of benchmark 
points has preferred somewhat those compatible with the present $g_\mu -
2$ measurement. 
Table~\ref{tab:1} shows the supersymmetric contribution to 
$a_\mu \equiv (g_\mu - 2)/2$, the relic density, and the
$B_{s\gamma}\equiv B(b\to s\gamma)$ for each benchmark point. 

The proposed points were not chosen to provide an `unbiased'
statistical sampling of the CMSSM parameter space but rather are intended to
illustrate different possibilities that are still allowed by the
present constraints~\cite{Battaglia:2001zp}, 
highlighting their different experimental 
signatures. Five of the chosen points are in the
`bulk' region at small $m_{1/2}$ and $m_0$, four are spread along the
coannihilation `tail' at larger $m_{1/2}$ for various values of
$\tan\beta$, two are in the `focus-point' region at large $m_0$, and two
are in rapid-annihilation `funnels' at large $m_{1/2}$ and $m_0$. Furthermore, 
the proposed points range over the allowed values of $\tan\beta$ from 5 up
to 35 and
50.  Most of the points have $\mu > 0$, as favoured by $g_\mu - 2$, but
there are also 
two points with $\mu < 0$.

\begin{figure}[t]
\includegraphics[scale=0.90]{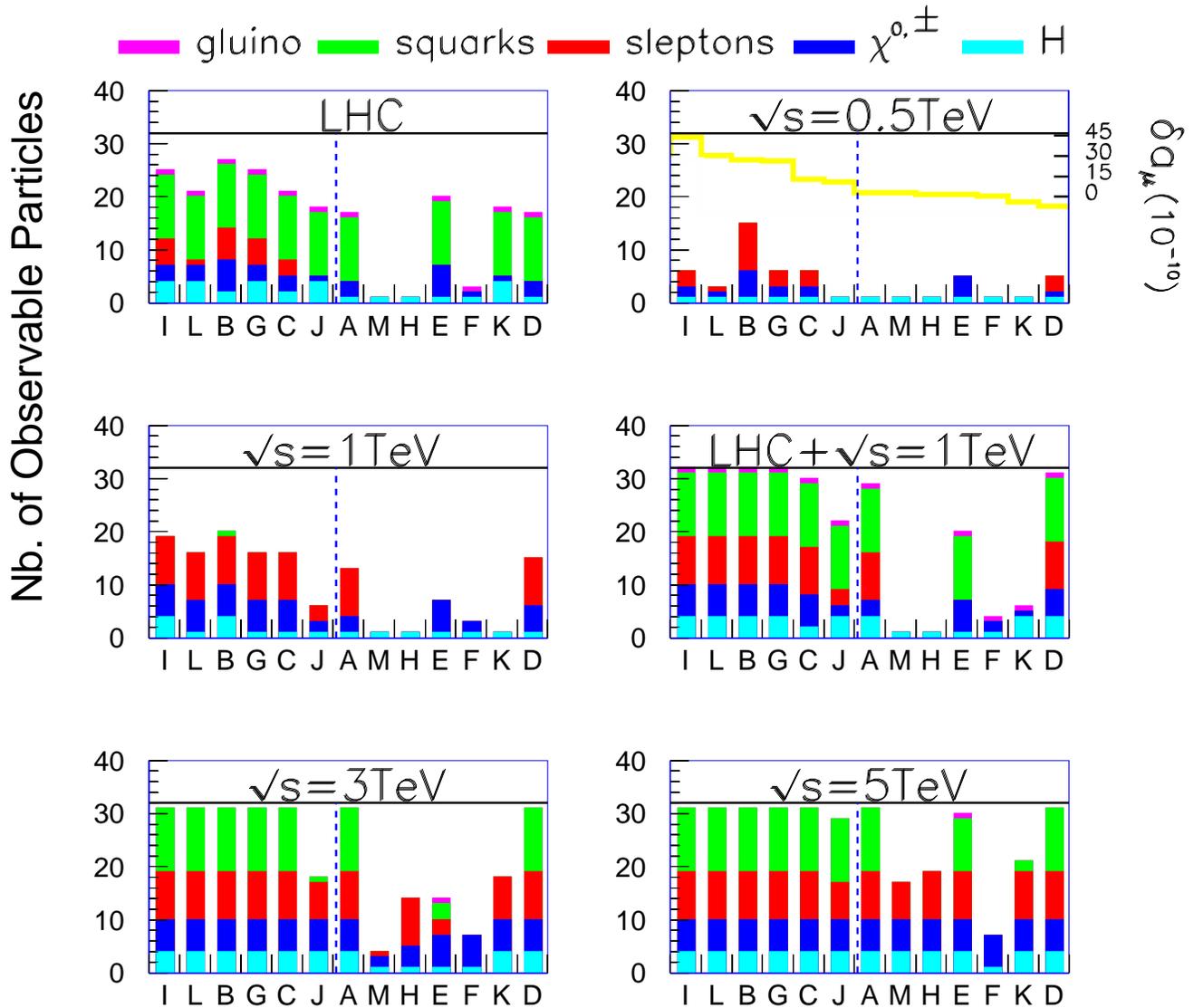}
\caption{Summary of the prospective sensitivities of the LHC, 
linear
colliders at 
different $\sqrt{s}$ energies and their combination in the 
proposed benchmark scenarios, which are
ordered by their distance from the central value of $g_\mu - 2$, as
indicated by the pale (yellow) line in the second panel. We see clearly
the complementarity
between an $e^+ e^-$ collider and the LHC in the TeV range of
energies~\cite{Battaglia:2001zp}, with the former
excelling for non-strongly-interacting particles, and the LHC for
strongly-interacting sparticles and their cascade decays. CLIC provides
unparallelled physics reach for non-strongly-interacting sparticles,
extending beyond the TeV scale. We
recall that mass and coupling measurements at $e^+ e^-$ colliders
are usually much cleaner and more precise than at
hadron-hadron colliders such as the LHC. Note, in particular, that it is
not known how to distinguish the light squark flavours at the LHC.}
\label{fig:2}
\end{figure}

\section{Discussion}

With time, some of the points we propose will become obsolete, for example
because of Higgs or SUSY searches at the Tevatron 
or reductions in the error in $g_\mu - 2$.
If there is no convincing indirect signal of new physics in low-energy 
experiments, the points in the coannihilation `tail', 
especially at its extreme tip, in the `focus-point' region and in
the rapid-annihilation `funnels' will be more difficult to exclude or
explore by direct detection. Some of these points might appear
disfavoured by fine-tuning arguments, but they cannot be excluded. Taken
together, 
the points proposed exemplify the range of different possible scenarios
with which 
future colliders may be confronted, and should provide helpful aids for
understanding 
better the complementarity of different accelerators in the TeV energy range. 

The physics reaches of various TeV-scale colliders: the LHC, a 500-GeV to 1-TeV linear 
$e^+ e^-$ collider such as TESLA, the NLC or the JLC, and a 3- to 5-TeV linear $e^+e^-$ 
collider such as CLIC have been estimated. The
detectability criteria adopted for the LHC are
discussed in detail in~\cite{Battaglia:2001zp}. 
For $e^+e^-$ colliders, the observability 
of each sparticle has been assessed on the basis of a required 0.1 fb for
the product of production cross section $\times$ observable decay branching 
fraction~\cite{Battaglia:2001zp}.
A grand summary of the reaches of the various accelerators is presented graphically in
Fig.~\ref{fig:2}. The different levels of shading (colour) present
the different types of sparticle: Higgses, charginos and neutralinos,
sleptons, squarks and gluino.  The first six points (I, L, B, G, C, J) are
presently favoured: they are compatible within 2 $\sigma$ with the present
$g_\mu - 2$ measurement, and the fine tuning is relatively small for most
of these points. Figure~\ref{fig:2} summarises the discussion
of~\cite{Battaglia:2001zp}, and exposes clearly the complementarity of
hadron and electron machines. It is apparent that many alternative 
scenarios need to be kept in mind. 

The LHC is expected to observe at least one CMSSM Higgs boson in all
possible scenarios, and will in addition discover supersymmetry in most of
the models studied. However, we do observe that the discovery of
supersymmetry at the LHC is apparently not guaranteed, as exemplified by
benchmarks H and M. It would be valuable to explore the extent to which
precision measurements at the LHC could find indirect evidence for new physics 
in such scenarios. We have chosen points at different values of $\tan \beta$, five 
of which are at large values, which may assist the LHC experiments in
assessing
the implications of the underlying phenomenology in the trigger 
and reconstruction of events. Some points, such as B and those at high $\tan\beta$, 
have final states rich in $\tau$s, point H involves a heavy long-lived
$\tilde{\tau_1}$,
and the different mass hierarchies between squarks and the gluino affect
the 
transverse energies and jet multiplicities of signal events. The CMS
Collaboration has 
started an investigation of the B, C, E and G benchmarks,
representative of 
these different scenarios, and analogous studies are foreseen by ATLAS.  
The need for high $\tan\beta$ points for LHC studies is dictated e.g. by the
experimentally 
challenging $H\rightarrow \tau\tau$ decays, for a Higgs with a mass
in the range of  300-500 Ge; this can be studied  with points I and L.

An $e^+ e^-$ linear collider in the TeV range would in most cases bring
important additional discoveries, exceptions being benchmarks H and M, and
possibly E. Moreover, such a linear collider would also provide many
high-precision measurements of the Higgs boson and supersymmetric particle
masses and decay modes, that would play a pivotal r\^ole in first
checking the CMSSM assumptions and subsequently pinning down its
parameters. In particular point B is a prime candidate to be studied
at such a collider.

In many of the scenarios proposed, the discovery and detailed measurements of the
complete set of supersymmetric particles, and especially some of the heavy
Higgses, gauginos and sleptons, will have to await the advent of a machine
like CLIC. For some of the proposed points, CLIC may even need to run at an energy 
considerably higher than 3~TeV. Distinguishing the different squark flavours could be 
an interesting challenge for CLIC. The CLIC potential in mapping the sparticle 
properties is presently being studied for points C, E and H.

\section{Prospects}

Our preliminary observations need now to be confirmed by more detailed
exploration of these benchmark scenarios. Moreover, we have not considered
benchmarks for models with gauge-mediated
\cite{Dine:1993yw,Dine:1995vc,Dine:1996ag}, gaugino-mediated
\cite{Kaplan:1999ac,Chacko:1999mi} or anomaly-mediated
\cite{Randall:1998uk,Giudice:1998xp} supersymmetry breaking, or models
with broken $R$ parity. Studies of additional benchmarks in these and
other models would represent interesting complements to this work. 
History reminds us that benchmarks have a limited shelf-life: at most one
of them can be correct, and most probably none. In future, the CMSSM
parameter space will be coming under increasing pressure from improved
measurements of $g_\mu - 2$, assuming that the present theoretical
understanding can also be improved, and $b \rightarrow s \gamma$, where
the $B$ factories will soon be dominating the measurements.  We also
anticipate significant improvement in the sensitivity of searches for
supersymmetric dark matter~\cite{Ellis:2001hv}. This may stimulate the
further redefinition of benchmarks for supersymmetry. However, we hope
that the diversity of sparticle spectra and experimental signatures
represented in these benchmarks will guarantee some general
validity for the conclusions that can be obtained from their detailed
study. 

\begin{acknowledgments}
\noindent
The work of K.A.O. was supported partly by DOE grant DE--FG02--94ER--40823.
\end{acknowledgments}

\bibliography{p3_47}

\end{document}